\documentclass[num-refs]{wiley-article}

\usepackage{siunitx}

\papertype{Original Article}

\title{A Systematic Review of Technical Defenses Against Software-Based Cheating in Online Multiplayer Games}

\author[1\authfn{1}]{Adwa Alangari}
\author[1\authfn{1}]{Ohoud Alharbi}

\contrib[\authfn{1}]{Equally contributing authors.}

\affil[1]{King Saud University}

\corraddress{Adwa Alangari, King Saud University}
\corremail{447205504@student.ksu.edu.sa}

\fundinginfo{The authors received no external funding for this research.}

\runningauthor{Adwa Alangari}

\begin{document}

\begin{frontmatter}
\maketitle

\begin{abstract}
This systematic literature review surveys technical defenses against software-based cheating in online multiplayer games. Categorizing existing approaches into server-side detection, client-side anti-tamper, kernel-level anti-cheat drivers, and hardware-assisted TEEs. Each category is evaluated in terms of detection effectiveness, performance overhead, privacy impact, and scalability. The analysis highlights key trade-offs, particularly between the high visibility of kernel-level solutions and their privacy and stability risks, versus the low intrusiveness but limited insight of server-side methods. Overall, the review emphasizes the ongoing arms race with cheaters and the need for robust, adversary-resistant anti-cheat designs.

\end{abstract}
\end{frontmatter}

\section{Introduction}
The global landscape of online multiplayer gaming has expanded into a multi-billion dollar industry, with competitive integrity serving as a foundational pillar of its success. However, this growth is continually threatened by software-based cheating, which undermines fair competition, degrades the player experience, and leads to significant economic losses for game developers due to player attrition. The problem is exacerbated by the client-server architecture of most games, where the client\textquotesingle s execution environment is fundamentally untrusted, allowing malicious actors to manipulate game state, inputs, and rendered output for an unfair advantage. Despite the critical nature of this problem, the technical literature on
anti-cheat mechanisms is fragmented, often focusing on single-solution prototypes or commercial systems with limited public disclosure. A comprehensive, comparative evaluation of the diverse technical defense categories---from non-invasive server-side models to highly privileged kernel-level drivers---is currently lacking. This gap makes it challenging for researchers to identify promising avenues for future work and for practitioners to make informed decisions regarding the deployment and trade-offs of different anti-cheat strategies. To address this, this Systematic Literature Review (SLR) aims to synthesize and evaluate the current state of technical defenses against software-based cheating. The findings are structured to provide a clear, comparative map of the strengths, weaknesses, and constraints of modern anti-cheat methodologies. To guide this systematic literature review, we define the following
research questions:

\begin{enumerate}
\def\labelenumi{\arabic{enumi}.}
\item
  What types of technical defenses have been proposed or implemented to
  detect or prevent software-based cheating in online multiplayer games?
\item
  How are these defenses empirically evaluated in terms of detection
  effectiveness, performance overhead, privacy impact, and scalability?
\item
  What are the key tradeoffs and constraints (e.g., latency, privacy,
  attacker adaptation, data requirements) observed across the different
  defense categories?
\end{enumerate}

These research questions frame the review, helping to systematically
extract, compare, and synthesize findings from prior work to provide
actionable insights for both researchers and practitioners.

\section{Related Work}
This section defines the key terms we use for the systematic review and
summarizes prior technical work that defends against software-based
cheating in online multiplayer games, organized by the defense
categories used in this study.

\subsection{Foundations for clear
definitions}\label{foundations-for-clear-definitions}
In this review the core definitional aim is to make explicit what we mean by software-based
cheating and which classes of technical defense we include under the
heading ``detect or prevent software-based cheating.''

This review adopt ``software-based cheating'' to mean automated or programmatic
manipulation of the client, its inputs, or the rendered output that
gives the actor an unfair advantage in multiplayer play; this includes
bots and input-automation (TAS/auto-players), binary patches and
injected code that expose or modify secret game state, and visual
overlays that disclose hidden information (e.g., ESP/wallhacks).

This review group technical defenses into five categories so that the review is
comparable across studies: (a) server-side detection (statistical/ML on
inputs, replays, or game logs), (b) client-side anti-tamper and
obfuscation (code packing, memory obfuscation, hash checks), (c)
kernel-level anti-cheat drivers, (d) hardware-assisted enclave/TEE
solutions (e.g., Intel SGX, ARM TrustZone) that move sensitive
data/logic into a protected execution domain.

These categories map to distinguish threat models and deployment
constraints (privacy, latency, trust), so structuring the SLR around
them makes it possible to compare defenses by what they protect and how
they change the attacker's cost/ability.

\subsection{Inclusion and Exclusion Criteria}
\label{inclusion-and-exclusion-criteria}

\subsubsection{Publication Type}
\label{publication-type}

Peer-reviewed technical papers published in reputable conferences and
journals were included. Peer-reviewed academic outputs provide
reproducible methods, implementation details, and performance
measurements necessary to compare defenses and reproduce experiments.

\subsubsection{Language}\label{language}

English (or English translation available); The corpus and evaluations
in the available documents are in English, review teams commonly
restrict to English for consistent interpretation unless multilingual
review capacity exists.

\subsubsection{Time period}\label{time-period}

This review includes studies published from 2019 onward, capturing the
period when modern anti-cheat approaches, machine learning--based
detection, and Trusted Execution Environments (TEEs) became prominent.
Focusing on this timeframe ensures the synthesis reflects current
technologies and practical deployment trends.

\subsubsection{Domain relevance}\label{domain-relevance}

ensures that included studies focus on software engineering aspects of
online multiplayer game security, particularly cheating detection,
client-side tampering, game-related TEEs (Trusted Execution
Environments), or kernel-level integrity defenses. Only work directly
applicable to gaming is considered, keeping the review focused on
solutions aligned with the software design, performance, and fairness
requirements of multiplayer systems.

\subsection{Search Strategy}\label{search-strategy}

\subsubsection{Keywords and Synonyms}
\label{keywords-and-synonyms}

The search strategy employed a comprehensive set of keywords and
synonyms related to online multiplayer games and software-based
anti-cheat mechanisms. Core terms included \emph{multiplayer games},
\emph{anti-cheat}, \emph{game security}, \emph{game integrity},
\emph{tamper detection}, \emph{memory editing}, \emph{binary
modification}, and \emph{software protection}. Synonyms and related
variants were incorporated to maximize coverage across different
terminologies used in the literature.

\subsubsection{Boolean Search Strings}
\label{boolean-search-strings}

The following Boolean search string was used across digital libraries
to retrieve relevant studies:

\begin{verbatim}
TS = (
  ("game*" OR "video game*" OR "online game*" OR "multiplayer game*" 
   OR "computer game*" OR "mobile game*" OR esports)
  AND
  ("anti-cheat*" OR "anti cheat*" OR "game security" OR "game integrity"
   OR "cheat detection" OR "cheating detection" OR "tamper detection"
   OR "anti-tamper" OR "client integrity" OR "cheat prevention"
   OR "game hacking" OR "game cheat*" OR "memory editing"
   OR "code injection" OR "reverse engineering"
   OR "man-at-the-end attack*" OR MATE
   OR "software tampering" OR "application tampering"
   OR "binary modification" OR "software protection")
)
NOT TS = (
  "academic cheating" OR "exam cheating" OR "student cheating"
  OR classroom OR education OR psychology OR "social cheating"
)
\end{verbatim}

\subsubsection{Databases and Digital Libraries}
\label{databases-and-digital-libraries}

To ensure comprehensive coverage of relevant literature, this review
selected major peer-reviewed databases and digital libraries commonly
used in software engineering and computer security research. These
sources provide broad disciplinary coverage, high-quality publication
standards, and robust citation indexing.

\begin{table}[htbp]
\caption{Databases and digital libraries used in the study}
\label{tab:databases}
\centering
\begin{tabular}{p{0.28\linewidth} p{0.62\linewidth}}
\hline
\textbf{Database} & \textbf{Justification} \\
\hline
IEEE Xplore &
Strong coverage of software engineering, computer security, and applied
machine learning research, including technical studies on anti-cheat
and game protection systems. \\ \hline

ACM Digital Library &
Comprehensive source for computing research, interactive systems, and
client-side security mechanisms relevant to online multiplayer games. \\ \hline

Scopus &
Broad multidisciplinary indexing database used to ensure completeness
of literature retrieval and support citation tracking. \\ \hline

Web of Science &
High-quality peer-reviewed coverage with strong citation indexing and
multidisciplinary reach. \\ \hline

ScienceDirect &
Extensive collection of journals in applied software engineering and
computer security, including research related to game security. \\ \hline
\end{tabular}
\end{table}
\clearpage

\subsection{Screening}\label{screening-1}

A comprehensive search was conducted in the Web of Science database
using a single search string. The screening process followed a
structured, multi-step workflow aligned with the PRISMA 2020 framework
to ensure transparency and reproducibility.

\begin{figure}[htbp]
\centering
\includegraphics{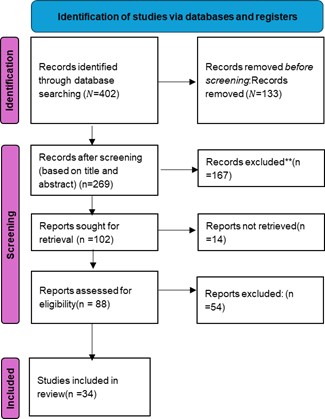}
\caption{Prisma 2020 framework}
\end{figure}

A total of 402 records were identified, of which 133 were removed before
screening. The remaining 269 records were screened based on titles and
abstracts, resulting in the exclusion of 167 records. Subsequently, 102
reports were sought for full-text retrieval, and 14 of these could not
be accessed. The remaining 88 full-text articles were assessed for
eligibility, and all were included in the final review. The screening
was conducted in three phases title screening, abstract screening, and
full-text screening to systematically narrow down relevant studies.
Rayyan, an online systematic review management tool, was used to
facilitate the screening process. Rayyan was chosen for its ability to
support blinded screening, tagging, and efficient organization of
records, ensuring an unbiased and well-documented selection process.

\section{Findings}\label{findings}

This section reports the results of the systematic literature review

\subsection{Synthesizing a definition for
anti-cheats}\label{synthesizing-a-definition-for-anti-cheats}

An anti-cheat system is a multi-layered technical defense mechanism
deployed in online multiplayer environments designed to maintain the
integrity of competition and the fairness of the user experience by
actively detecting, mitigating, or preventing unauthorized
modifications, exploits, or automated software (cheats) used by players
to gain an unfair. \cite{Brandao2020}, \cite{Spijkerman2020}
, \cite{PreventionDetection}

These systems are necessitated by the constant evolution of cheating
software, which exploits the core vulnerability of online gaming: the
player's control over the client-side execution environment \cite{Anwar2023}
.

The primary function of an anti-cheat is to guarantee that all
participants operate on an equal playing field by policing modifications
to the game code, configuration data, and execution environment \cite{Brandao2020}
.
This is achieved through two complementary strategic goals. Integrity
enforcement (Prevention) proactively securing game state data and code
against tampering, often through architectural design, code hardening,
or hardware isolation to make cheats impossible \cite{PreventionDetection}
, \cite{Bauman2016}
. The
second goal is Malicious Behavior Detection (Reaction), identifying and
classifying suspicious activity that deviates from human norms or
legitimate parameters, leading to sanctions such as account bans
\cite{Alayed2013}
.

\subsubsection{Technical Implementations and
Methodologies}\label{technical-implementations-and-methodologies}

Client-side techniques focus on protecting the game environment and
detecting cheat software running on the player's machine, despite the
fundamental lack of trust in that. \cite{Dorner2024}

Kernel-Level Operations (Ring 0) methodology involves deploying
anti-cheat software as a kernel driver, operating at the highest
privilege level to monitor and restrict processes.\cite{Dorner2024}
Drivers intercept critical system calls frequently used by cheats (e.g.,
WriteProcessMemory, CreateRemoteThread) to redirect execution flow for
inspection and blocking via techniques like inline hooking.
Driver/Process Blocking makes the system prevents the game from
launching if unsigned drivers are detected, or if known
vulnerable/malicious kernel modules (like those used by Cheat Engine)
are loaded. Integrity Enforcement: The kernel module can continuously
scan kernel space and process memory for manually mapped DLLs, anomalies
(executable pages not backed by a proper module), or specific cheat
signatures. \cite{Heiser2016}
, \cite{Maario2021}

Software Obfuscation and Anti-Analysis are preventative measures
designed to increase the complexity and time required for attackers to
reverse-engineer the game \cite{Gjonbalaj2023}. Encrypting the game executable's code
section, sometimes decrypting functions only just before execution
(on-demand partial decryption) and re-encrypting immediately afterward,
to thwart static analysis. Also, there is memory obfuscation employing
techniques like value relocation (changing a variable's memory address
upon every modification) and encryption (e.g., XOR ciphers) to protect
dynamically allocated variables in the heap from memory scanners like
Cheat Engine. And Anti-Debugging for Actively checking for the presence
of a debugger attached to the game process or barraging hypervisor-based
debuggers with VM-Exits to reveal their presence. \cite{Kaiser2009}

Hardware-Assisted isolation methodologies leverage hardware features for
fundamental security guarantees, mitigating the untrusted nature of the
host. \cite{Bauman2016}
In a Intel SGX (Trusted Execution Environment) the cheat
detection engine or critical game logic is moved into an isolated
enclave to protect its integrity and confidentiality from the
operating system and other untrusted software. For example Trusted
Visibility Testing (BMTest) is used by systems like BlackMirror to
prevent wallhacks by performing secure, in-enclave visibility testing on
sensitive entities before selectively disclosing data to the untrusted
GPU.\cite{Park2020} It runs a runtime integrity check that stores hashes of
monitored game binaries are compared against runtime memory content to
detect modification of code or virtual function. \cite{Jeon2021}, \cite{Brandao2020}

Server-side methods offer a less invasive, scalable, and typically more
tamper-resistant approach by analyzing behavior and input validation.
Server-Authoritative design is the foundational methodology where the
server performs all crucial checks and simulations, refusing to blindly
trust client input. The server only sends data strictly necessary for
client rendering to minimize information exposure (e.g., preventing
wallhacks). \cite{Kang2013}

Machine learning and statistical anomaly detection which analyze vast
quantities of data to find patterns indicative of non-human behavior.
Behavioral modeling includes a training of deep learning models (e.g.,
RNN, CNN, LSTM, Transformers) on player data captured from game replays
or \cite{Pinto2021}
, \cite{Loo2025}
. Feature engineering and extracting metrics like
ViewAngleDifference, TrueViewAngle (recoil compensated), and Intersect
Point variance to detect aimbot snap and no-recoil patterns. \cite{Alkhalifa2016}
HCI data analysis where treating player input (keystrokes, mouse
dynamics) as multivariate time series to detect abnormal precision or
timing (e.g., for aimbot or triggerbot detection). \cite{Loo2025} Utilizing
log-mining techniques to identify statistical outliers at a group level.
For instance, analyzing party-play logs in MMORPGs based on metrics like
party duration, entropy index (diversity of actions), and action log
ratios (e.g., high `getting experience' vs.~low `quest completion').
\cite{Kang2013} Some also conduct a trajectory analysis employing manifold
learning and dissimilarity measures (like Markov Chain models or
Kullback--Leibler divergence) on avatar movement paths to distinguish
human variation from rigid bot scripts. \cite{Pao2010}

\subsection{How Defenses Are Empirically
Evaluated}\label{how-defenses-are-empirically-evaluated}

Empirical evaluation in the anti-cheat literature is not uniform; it
depends heavily on the defense category. Broadly, four dimensions recur:
detection effectiveness, performance overhead, privacy impact, and
scalability.

Detection effectiveness is measured most rigorously for behavioral and
ML-based defenses, which follow standard supervised-learning practice:
they define cheater/non-cheater (or specific cheat types), train on a
subset, and report accuracy, precision, recall, F1 and ROC-AUC on
held-out frames, sessions or players. In a large Mario Kart Wii study,
Weatherton's CatBoost models reach F1 = 0.7725 at the frame level using
both input and game-state features, while aggregating to full races
boosts F1 to 0.9109; after excluding anomalous developer ghosts the
session-level F1 climbs to 0.9583 with 1.000 precision on non-developer
players, i.e., no false positives in that subset. Deep time-series
models on HCI data show similarly strong discrimination: report 99.2\%
triggerbot and 98.9\% aimbot detection accuracy on CS:GO using CNNs on
mouse/keyboard streams, and a Minecraft system trained on 40 minutes of
normal vs aimbot/automining mouse data from 13 users achieves F-scores
above 99\% with both 2D-CNN and LSTM architectures. Earlier
trajectory-based work finds Quake-style bots can be detected with
\textgreater95\% accuracy once 200+ seconds of avatar movement are
observed, with error and false-positive rates falling as the observation
window grows. \cite{Pinto2021}
, \cite{Pao2010}
, \cite{Weatherton2025}
, \cite{Lukas2022}

Protocol and architecture-level defenses are judged by whether they make
whole exploit classes impossible or reliably detectable. In Lehtonen's
number-guessing game, the naïve design lets the client compute battle
outcomes and simply report wins, so a cheater can succeed by modifying
or replaying a single packet; after refactoring to a
server-authoritative model that recomputes and verifies state, the same
packet-tampering no longer affects results. This before/after experiment
underpins his rating of ``not trusting the client'' and tamper-resistant
protocols as maximally resistant to tampering (4/4) in his comparative
table. \cite{Lehtonen2020}

Commercial stacks and trusted-execution approaches are evaluated with
targeted attack suites and concrete cheats. grey-box attacks (user- and
kernel-mode injection, code patching, debug attachment, hidden drivers,
etc.) against 11 popular FPS/Battle Royale titles, recording whether
each is proactively blocked, reactively banned, merely crashes the game,
or goes undetected; Valorant (Vanguard) and Fortnite (EAC+BattlEye)
which considered kernel-level anti-cheat drivers, though they also
incorporate server-side detection methods. are the only configurations
that consistently prevent or detect most tests, whereas purely user-mode
setups like classic VAC or FairFight-only allow many kernel-mode or
early-boot attacks to succeed. \cite{Collins2024} SGX-based prototypes demonstrate
similar ``all-or-nothing'' effectiveness for their targeted threats: an
SGX-hardened CS:GO prototype moves the detection engine and critical
configuration into enclaves plus a kernel module and successfully
detects and blocks the open-source Fuzion cheat, while an unprotected
baseline is trivially compromised, \cite{Brandao2020}
 and BlackMirror shows that
once Quake II's visibility logic and entity states are confined to SGX
and only clipped visibility is exposed to the untrusted GPU, standard
wallhack techniques (GDI hooking, overlay rendering) can no longer
reveal enemies through walls.\cite{Park2020} Finally, large-scale industry
systems are evaluated via operational outcomes rather than public ML
metrics. FairFight, used in Battlefield V and Titanfall 2, is described
as effective at catching statistical outliers in server telemetry,
though its exact precision/recall figures remain proprietary. \cite{Kang2013}, \cite{Weatherton2025}

Performance overhead is measured via both precise micro-benchmarks and
coarse scoring. SGX-based systems also provide empirical overhead
measurements. Although BlackMirror primarily emphasizes security, its
Quake II implementation reports that moving visibility computations into
enclaves adds measurable but tolerable latency, keeping frame rates
within real-time bounds and maintaining smooth gameplay on tested maps.
\cite{Park2020}, \cite{Brandao2020}

Kernel-level integrity enforcement like Kenali (data-flow integrity for
Linux kernels) shows that focusing on a small inferred set of critical
kernel structures can reduce overhead to 1.5\%--3.5\% on common
benchmarks. While not an anti-cheat, it suggests that carefully targeted
kernel checks can be deployed without prohibitive performance loss. \cite{Kaiser2009}

Detailed benchmarks are not provided, studies rely on relative scoring.
Lehtonen rates each method 1--4 for ``lack of overhead.'' Server-side
statistical methods and SaaS-style ML systems working on logs are scored
4 (negligible overhead) because they run out-of-band on backend
infrastructure. File hashing and signature scanning are likewise scored
4 because they are typically performed at install or patch time. In
contrast, ``do not trust the client'' designs and complex
tamper-resistant protocols receive they require the server to simulate
or validate every action, increasing CPU use and network round-trips,
which can negatively affect latency in fast-paced games if
infrastructure is under-provisioned. Partial code encryption and
aggressive memory obfuscation also receive a 2 due to the cost of
frequent decryption/encryption of hot code and heap objects. \cite{Lehtonen2020}

\begin{table}[htbp]
\caption{Lehtonen framework evaluation of anti-cheat methods}
\label{tab:lehtonen-framework}
\centering
\small
\begin{tabular}{p{0.18\linewidth} p{0.12\linewidth} p{0.15\linewidth} p{0.10\linewidth} p{0.12\linewidth} p{0.18\linewidth}}
\hline
\textbf{Method} & \textbf{Resistance to Tampering} & \textbf{Ease of Implementation} & \textbf{Lack of Overhead} & \textbf{Non-invasiveness} & \textbf{Suitability for Wide Variety of Games} \\
\hline
\textbf{Server-side methods} & & & & & \\
Not trusting the client & 4 & 2 & 2 & 4 & 3 \\
Tampering resistant application protocol & 4 & 2 & 2 & 4 & 4 \\
Obfuscating the network traffic & 2 & 4 & 3 & 4 & 4 \\
Statistical methods & 3 & 1 & 4 & 4 & 2 \\
\textbf{Client-side methods} & & & & & \\
Code encryption & 2 & 1 & 2 & 4 & 4 \\
Verifying files by hashing & 1 & 4 & 4 & 3 & 4 \\
Detecting known cheat programs & 1 & 2 & 4 & 1 & 4 \\
Obfuscating memory & 2 & 2 & 2 & 4 & 4 \\
Kernel-based anti-cheat driver & 3 & 2 & 4 & 2 & 4 \\
\hline
\end{tabular}
\end{table}

Privacy impact is generally evaluated through architectural and legal
analysis rather than quantitative metrics. Kernel-level anti-cheat
drivers are rated 2/4 on non-invasiveness in Lehtonen's framework. They
run at ring-0 with full access to all user-space memory and OS
structures. Dorner and Klausner's rootkit-oriented evaluation of
BattlEye, Easy Anti-Cheat, FACEIT AC and Vanguard finds clear
rootkit-like properties in the latter two: load-on-boot kernel drivers,
aggressive self-protection, extensive scanning of drivers and processes,
mandatory Secure Boot or hypervisor settings, and continuous logging of
hardware identifiers. \cite{Dorner2024}

Greidanus's legal analysis of client-side anti-cheat under EU ePrivacy
and GDPR reinforces this view. He argues that scanning beyond the game's
own process (e.g., probing other processes, DNS caches, or taking
screenshots) generally falls outside the ``strictly necessary''
exception of Article 5(3) of the ePrivacy Directive and thus requires
informed consent. In practice, he finds that most game privacy policies
and EULAs mention anti-cheat only vaguely (``monitor RAM,'' ``detect
unauthorized third-party software'') and do not disclose kernel-mode
loading, cross-process scanning, or potential screenshot capture. This
lack of transparency, combined with the breadth of access, underlines
his conclusion that many client-side engines are functionally equivalent
to spyware or rootkits in legal terms. \cite{Greidanus}

Scalability is assessed in two ways implementation effort and
portability across games, and operational behavior in large-scale
deployments.

From an implementation perspective, scalability is divided into the ease
of integration and the suitability for a variety of games. Simple,
game-agnostic techniques like network encryption and file hashing are
highly scalable in this regard, scoring high marks (4/4 for suitability
and ease) because they rely on mature libraries and straightforward
integration across multiple titles without needing specialized domain
knowledge. Conversely, statistical and machine learning (ML) anti-cheat
methods exhibit poor implementation scalability, scoring low for ease
and suitability. \cite{Oster2025}
 \cite{Oster2025}

Operational scalability, the second dimension, is robustly demonstrated
by cloud-based and platform-level systems. Services like FairFight and
Easy Anti-Cheat (EAC) are offered as scalable, centralized SaaS
products, a model proven effective by their adoption in major titles
like Battlefield V, Titanfall 2, and Fortnite. These systems efficiently
ingest massive volumes of telemetry from many sources and apply
configurable, data-driven rules at scale, often requiring significant,
centralized processing power, as seen with Valve's VACNet, which
processes hundreds of thousands of matches daily. \cite{Gjonbalaj2023}

\subsection{Key Tradeoffs and Constraints Across Defense
Categories}\label{key-tradeoffs-and-constraints-across-defense-categories}

The evaluations above reveal a set of recurring tradeoffs and
constraints that shape practical anti-cheat design.

The first tension is privacy vs observational power in client-side
defenses. Kernel-mode anti-cheats and cross-process scanners are
empirically the most effective for thwarting injection, patching and
kernel-level cheats, but they achieve this precisely because they
operate with rootkit-like privileges. This creates clear privacy and
security risks: they can destabilize systems, conflict with drivers and
security tools, expose highly sensitive data if compromised, and are
difficult to reconcile with data-protection principles. \cite{Dorner2024}
, \cite{Maario2021}

\begin{table}[htbp]
\caption{Kernel vs User Space}
\label{tab:kernel-vs-user}
\centering
\small
\begin{tabular}{p{0.18\linewidth} p{0.36\linewidth} p{0.36\linewidth}}
\hline
\textbf{Feature} & \textbf{User Space} & \textbf{Kernel Space} \\
\hline
Memory access & Limited access & Full access \\
Hardware access & No direct access & Full access \\
Access to CPU instructions & Only unprivileged instructions & All instructions \\
Access to critical OS data structures & No access & Full access \\
\hline
\end{tabular}
\end{table}

Server-side defenses occupy the opposite end of this spectrum. They are
inherently privacy-friendly and easier to justify legally, but their
visibility is limited to what clients send over the network. They cannot
directly see overlays, wallhacks that only affect rendering, or external
vision-based aimbots running on separate machines. Their detection power
is confined to statistical anomalies in movement, accuracy, timing, or
network-level behavior. Latency and responsiveness impose another
constraint. Strict ``do not trust the client'' designs and lockstep
protocols improve integrity by ensuring the server or a referee
validates all state changes before they take effect, but this increases
server load and network round-trips. In high-tick-rate FPS or fighting
games, even small latency additions are perceptible and can degrade the
experience. Many commercial architectures therefore adopt partial trust
and prediction (client-side prediction, server reconciliation) to
balance security and responsiveness. \cite{Oster2025}

Machine-learning-based behavioral detection introduces its own
tradeoffs. High detection performance in research often relies on large,
high-quality labeled datasets (replays, input logs, network traces),
stable cheat behaviors (e.g., consistent aimbot snap patterns), and
relatively stationary game mechanics \cite{Joens2024}
 \cite{Weatherton2025}

In production, acquiring labels at scale is expensive and noisy (e.g.,
relying on player reports). Cheats and player strategies evolve, causing
concept drift; models must be retrained and monitored to avoid
performance degradation. Attackers can also attempt to adapt, e.g., by
making aimbots mimic human jitter or by injecting random ``mistakes,''
as shown in GAN-Aimbots and later work. This arms-race dynamic forces
defenders into continuous iteration. \cite{Kanervisto2023}

Data requirements and computing constraints further limit who can
realistically deploy heavy ML. VACNet-style architectures with deep
models, or large-scale sequence models, are largely confined to platform
holders or AAA studios that can afford to log vast amounts of gameplay
and maintain substantial compute clusters. Smaller studios often cannot,
and either rely on generic third-party SaaS anti-cheats or fall back on
simpler rule-based systems. \cite{Jonnalagadda2021}

Hardware-assisted defenses (SGX, TrustZone) offer strong integrity
guarantees but introduce platform and maintainability constraints. SGX
is not ubiquitous on consumer desktops, has its own side-channel issues,
and requires developers to partition code carefully into enclaves.
TrustZone is common on mobile SoCs but demands device-vendor cooperation
and per-title integration. Both approaches are therefore more suitable
for flagship titles or platforms with tight hardware--software
integration than for the broader market. \cite{Park2020}, \cite{Brandao2020}

Finally, attacker adaptation and cheat economics interact with all
categories. Collins et al.~show that stronger kernel-level anti-cheats
are correlated with higher cheat subscription prices and lower observed
cheat uptime, indicating that robust defenses do not eradicate cheating
but increase its cost and operational friction. \cite{Collins2024}

\section{Discussion}

This systematic literature review also expands on simply cataloguing the
various anti-cheat technologies to include a critical synthesis of the
continued struggle between competitive integrity and adversarial
ingenuity. The results indicate that the current landscape is not one of
a single solution to cheating, but rather an intricate web of technical
trade-offs, ethical considerations and an escalating
\textquotesingle arms race\textquotesingle{} between game developers and
cheaters. The root challenge remains the basic asymmetry of trust: the
game client, executing on the player\textquotesingle s machine, is
always going to be considered untrustworthy; however, it is the only
element of the game that has the required proximity to the game state to
sufficiently police it. Therefore, my discussion provides a structured
examination of the reviews' most important contributions.\cite{Wang2019}

\subsection{Key Findings}

My synthesis clearly indicates a division in the philosophy of
anti-cheating solutions based on the degree of system access and the
point of intrusion. The most important finding was the emergence of a
dual-pronged defensive approach.

The invasive, high-privileged kernel level driver (i.e., Vanguard)
provides the greatest observation and the best active defense against
sophisticated cheating. Its ability to intercept and control the system
calls relied upon by cheats to execute them makes it nearly impossible
for sophisticated attackers to gain entry.\cite{Dorner2024}

Non-invasive, server-side behavioral models utilizing deep learning
provide a privacy preserving, scalable and passive layer of defense. As
opposed to detecting how a cheat executes its mechanism, these models
detect the consequences of cheating and therefore provide a tamper proof
layer of defense isolated from client-side bypasses.\cite{Zhan}

Hardware Assisted Enclaves (HAE) such as Intel SGX represent an
unfulfilled promise. Although HAEs provide a secure, isolated
environment for critical logic, they have not been able to be
practically deployed due to high complexity and performance overhead,
along with the fact that they are dependent on hardware platforms. HAEs
can represent a powerful, targeted solution for certain types of threats
(e.g., wallhacks); however, they do not represent a solution to all
types of cheating.\cite{Alayed2013}
\cite{Heiser2016}

\subsection{Synthesis of Trade-offs and Constraints}

Deployment of anti-cheating solutions is a constrained optimization
problem. An effective anti-cheating solution will require a layered
defense that must manage inherent trade-offs in four critical areas:
effectiveness, performance, privacy and scalability.

\begin{table}[htbp]
\caption{Findings of defense categories}
\label{tab:defense-findings}
\centering
\small
\begin{tabular}{>{\raggedright\arraybackslash}p{0.18\linewidth} 
                >{\raggedright\arraybackslash}p{0.13\linewidth} 
                >{\raggedright\arraybackslash}p{0.09\linewidth} 
                >{\raggedright\arraybackslash}p{0.12\linewidth} 
                >{\raggedright\arraybackslash}p{0.12\linewidth} 
                >{\raggedright\arraybackslash}p{0.20\linewidth}} 
\hline
\textbf{Defense Category} & \textbf{Detection Effectiveness} & \textbf{Performance Overhead} & \textbf{Privacy / Stability Risk} & \textbf{Scalability / Deployment} & \textbf{Key Constraint} \\
\hline
Server-Side (ML/Behavioral) & High (Reactive) & Low & Very Low & Excellent & Latency of Ban \\
Client-Side (Anti-Tamper) & Low (Easily Bypassed) & Low to Moderate & Low & Excellent & Untrusted Environment \\
Kernel-Level (Ring 0) & Very High (Proactive) & Moderate to High & Very High & Moderate (OS/Hardware dependency) & Ethical / Privacy Backlash \\
Hardware-Assisted (TEE) & High (Proactive / Targeted) & Moderate to High & Low (High implementation risk) & Low (Platform dependency) & Hardware Adoption Rate \\
\hline
\end{tabular}
\end{table}

Perhaps the most acute constraint is the Privacy Effectiveness Paradox.
The greatest amount of effectiveness can be realized when using a kernel
level system. However, the greater the effectiveness, the higher the
system privileges required to realize that effectiveness. The higher the
system privileges required, the higher the risk to both user privacy and
system stability. On the other hand, anti-cheating solutions that
respect user privacy (server-side solutions) will be inherently limited
in their ability to prevent low-level, memory-based attacks. This
creates a difficult ethical and technical dilemma for developers who
must find a balance between the two competing interests. Additionally,
the Adaptation Constraint defines that any static or signature-based
solution will eventually reach its limits. Due to the nature of the
adversaries attempting to cheat, there will be a constant and resource
intensive need for updates to maintain effectiveness. In order to combat
this, there will be a fundamental shift away from signature-based and
static detection mechanisms toward behavioral and heuristic detection
mechanisms that are much harder for reverse engineers to exploit. The
definition of the arms race is the pace at which the adversaries adapt
to the defenses.\cite{Greidanus}

\subsection{Implications for Practice}

The findings of this study provide numerous actionable implications for
anti-cheating solution deployments by game developers and security
practitioners.

Firstly, the only viable method of deploying an anti-cheating solution
is a defense-in-depth strategy. The era of relying on a single
anti-cheating solution is over. The kernel level solution will act as
the primary deterrent and the server-side solution will act as the
ultimate, tamper-free arbiter.

Secondly, the development community should invest the majority of
resources in the advancement of features and deep learning models used
in server-side detection. Server-side detection models represent the
future of scalable, privacy conscious anti-cheating solutions that are
capable of identifying subtle, non-human patterns of play that bypass
simple signature checking. To accomplish this, developers must move
beyond simple metrics and implement complex multivariate time series
analysis of player input dynamics (Human Computer Interaction - HCI
data) and avatar trajectory analysis. By doing so, the developers will
create a unique digital fingerprint of authentic human play that will
help identify cheating activity.

Thirdly, security professionals must establish a policy of radical
transparency in regards to their anti-cheats operation, particularly
where kernel level access is concerned. Transparency in the form of
clear, publicly communicated policies and procedures, as well as
independently audited security practices, will help to mitigate the
risks associated with the use of kernel level access and help to foster
trust among players, which is equally as important as the technical
protection provided by the anti-cheat solution.

\subsection{Implications for Theory}

The results of this study challenge and expand existing theoretical
models in computer security. The traditional security boundary is
insufficient in the context of competitive online gaming. The work
presented here illustrates the need for a new theoretical model, the
Adversarial Trust Model - where the client is no longer simply
untrusted, but instead malicious. The design of the defense must take
place under the assumption that the client is hostile and that the
security concerns should not be focused on the integrity of the client
code, but instead the authenticity of the player\textquotesingle s
behavior.

Additionally, the trade-off limitations noted in the TEE literature
illustrate the need for additional theoretical work on Minimal Trust
Computing in gaming. Minimal Trust Computing in gaming refers to the
identification of the minimum number of game logic and data that must be
isolated to ensure competitive integrity. Minimal Trust Computing in
gaming is expected to minimize the attack surface and reduce the
complexity of deployment, resulting in the achievement of maximum
security with minimal system footprint.\cite{Bauman2016}

\subsection{Limitations and Outlook}

There are obvious limitations to this systematic review due to the
inherent secrecy of the anti-cheat industry. Much of the most advanced
and proprietary work done in the anti-cheat industry is not published in
peer-reviewed journals, resulting in a possible publication bias toward
academic prototypes and older commercial systems that may not be as
effective as some of the newer solutions. The authors\textquotesingle{}
analysis of kernel-level solutions, for example, is primarily based on
the limited public disclosure information and reverse-engineering
analyses available, rather than the actual first party technical
documentation. The lack of publicly disclosed information is a
significant limitation in obtaining a comprehensive view of the state of
the art.\cite{Alayed2013}

In terms of the outlook for the anti-cheat industry, the near future
will likely be characterized by two main factors. The first factor will
be the maturation of Federated Learning, which is expected to allow for
the collaborative, privacy-respectful sharing of behavioral cheat
signatures across multiple game titles, greatly increasing the speed of
detection of novel cheats. The second factor will be the growing
adoption of Cloud Gaming, which is expected to provide a revolutionary
architectural solution to the trust issue, by migrating the entire game
client execution environment to a trusted, remote server, thus
eliminating the client-side attack vector. While Cloud Gaming presents
many new challenges (such as increased streaming latency), it represents
the largest paradigmatic shift in the anti-cheat landscape to date, and
could end the \textquotesingle arms race\textquotesingle{} as we
currently understand it, by completely changing the security boundary.
The next generation of research should therefore focus on the security
and performance implications of this architectural revolution.\cite{Zhao2021}

\section{Conclusion}\label{conclusion}

This Systematic Literature Review has provided a comprehensive synthesis
of technical defenses against software-based cheating in online
multiplayer games, successfully categorizing the diverse landscape of
solutions and evaluating them across critical dimensions. By answering
the defined Research Questions, this work establishes a clear map of the
anti-cheat ecosystem, ranging from non-invasive server-side models to
high-privilege kernel drivers and hardware-assisted isolation.

The central conclusion is that effective anti-cheat strategy must be
multi-layered and adaptive. No single technology offers a complete
solution; instead, a combination of non-invasive behavioral detection
(server-side) and high-assurance integrity enforcement
(hardware-assisted TEEs) represents the most balanced and sustainable
path forward. The persistent tension between security effectiveness and
user privacy/stability, particularly evident with kernel-level
solutions, remains the most significant challenge. This review serves as
a foundational resource for both researchers seeking to identify
promising areas for innovation and practitioners tasked with deploying
robust, user-friendly anti-cheat systems in the face of an ever-evolving
threat landscape.


\begin{thebibliography}{99}
\bibitem{Zhan} ZHAN H, "Detection system for anti-cheating system, has double-coder for verifying monitoring microprocessor, where double-coder realizes preventing switching control main board cheating and inspecting cheating," doi:10.3390/CN110627011-A.
\end{thebibliography}

\end{document}